\documentclass[pre,aps,twocolumn,tightenlines]{revtex4}
\usepackage{times,graphicx,psfrag,amsmath,amsfonts}
\newcommand{\gpfig}[1]{%
  \resizebox{8cm}{!}{\includegraphics{#1}}\\[\baselineskip]
  }
\graphicspath{{eps/}}
\newcommand{\xlabel}[1]{\psfrag{xlabel}[][]{#1}}
\newcommand{\ylabel}[1]{\psfrag{ylabel}[b][t]{#1}}
\newcommand{\OrderOf}[1]{\ensuremath{\mathcal O}\left(#1\right)}
\DeclareMathAlphabet{\mathrmb}{OT1}{ptm}{b}{n}
\DeclareMathAlphabet{\mathsfb}{OT1}{phv}{b}{n}

\newcommand{\rmd}{{\mathrm{d}}}

\newcommand{\rmdu}{\rmd \Vector{u}}

\newcommand{\pder}[2]{\dfrac{\partial#1}{\partial#2}}
\newcommand{\Int}[3]{\int\limits_{#1}^{#2}\!\!{\rmd}#3\,}
\newcommand{\tInt}[3]{\int_{#1}^{#2}\!\!{\rmd}#3\,}

\newcommand{\Vector}[1]{\ensuremath{\mathrmb{#1}}}

\newcommand{\braket}[1]{\left\langle#1\right\rangle}
\newcommand{\dotpr}[2]{#1\!\cdot\!#2}

\newcommand{\evalat}[2]{\left.#1\right\vert_{#2}}
\newcommand{\Exp}[1]{{\mathrm{e}}^{#1}}
\newcommand\textfrac[2]{#1/#2}

\newcommand{\ie}{\textit{i.e.}\@}

\newcommand{\Figref}[1]{Fig.~\ref{#1}}
\newcommand{\Figsref}[1]{Figs.~\ref{#1}}
\newcommand{\Eqref}[1]{Eq.~\eqref{#1}}
\newcommand{\Eqsref}[1]{Eqs.~\eqref{#1}}
\newcommand{\secref}[1]{Sec.~\ref{#1}}

\newcommand\collc{\mathbb{C}}
\newcommand\hatf{\hat{f}}
\newcommand\rst{\rho^{*}}
\newcommand\scx{\sqrt{c}\,x}
\newcommand\epsl{\epsilon_{l}}
\newcommand{\ezfact}{\left(\tfrac{\epsl}{\sqrt{\pi}}\,\Exp{z}\right)}
\newcommand{\ndp}{N\sigma}
\newcommand{\rhonot}{\rho^{\mathrm{o}}}

\begin{document}

\title{An approximate solution to the Boltzmann equation for vibrated
granular disks}
\author{P Sunthar} 
\email{sunthar@chemeng.iisc.ernet.in}
\author{V Kumaran}
\email{kumaran@chemeng.iisc.ernet.in}

\affiliation{Department of Chemical Engineering,\\
             Indian Institute of Science, Bangalore, India}

\begin{abstract}
  The behaviour of the lower order moments of the velocity
  distribution function for a system of inelastic granular disks
  driven by vertical vibrations is studied using a kinetic theory.  A
  perturbative kinetic theory for vibro-fluidised beds was proposed by
  Kumaran (JFM, v. 364, 163). A scheme to generalise this theory to
  higher orders in the moments is presented here.  With such a method
  it is possible to obtain an analytical solution to the moments of
  the distribution function up to third order.
  \end{abstract}

\maketitle

\section{Introduction}

The dynamics of vibrated granular materials, its instabilities, and
pattern formation are of some interest in the recent years as
demonstrated by experiments of \cite{umbanetal96} and simulations of
\cite{ludetal94}.  The theoretical description of such systems is
complicated by the fact that it is a driven dissipative system
characterised by highly inelastic collisions and hence the validity of
equations of hydrodynamics is not clear at present \cite{kadanoff99}.
However, it is possible to describe one idealised situation, where the
dissipation due to particle collisions is small and the amplitude of
wall oscillations is small compared to the mean free path, as was
shown in the kinetic theories \cite{kum98:vib,sunkum99:scal}.  Such a
description might be one of the starting points where we can ascertain
with some confidence the rigour of the approach used. The present work
is a continuation of such an approach.

In this communication, we show that it is possible to obtain an
analytical solution to the Boltzmann equation for a dilute bed of
vibrated granular disks, by the method of moments, correct upto third
order in the moments of the distribution function.  The method of approach
followed holds the same principle as followed in \cite{kum98:vib},
except in the choice of the distribution function, which is done here
by expanding it in the orthogonal set of Hermite polynomials.  It is
hoped that the analysis of this base state solution for its stability
would give some clues to understanding the instabilities occuring in a
vibro fluidised bed.

In \secref{sec:form}, we present a general methodology to approximate
the distribution function by expanding it in Hermite polynomials, and
a procedure to solve the Boltzmann equation. We present only the
important results here, the reader is referred to \cite{kum98:vib} for
the details of the kinetic theory of vibro-fluidised beds. In
\secref{sec:third}, we obtain an analytical solution when the
formulation is restricted upto the third order in the moments. The results
we obtain here are qualitatively similar to results of \cite{kum98:vib},
except that we have obtained an analytical solution whereas in the latter
it was an approximate series solution.

\section{Basic formulation} \label{sec:form}
The Boltzmann equation for the velocity distribution function,
$f(\Vector{x},\Vector{u})$, for vertically vibrated beds is \cite{kum98:vib}:
\begin{equation}
  \label{eq:beqn}
 \partial_{t} \hatf + u_{i}^{*} \partial_{i^*} \hatf - g \pder{\hatf}{u_z^{*}}
   = \frac{\partial_{c} \hatf}{\partial t}
\end{equation}
where the collision integral is,
\begin{equation}
  \label{eq:colldef}
  \frac{\partial_{c} \hatf}{\partial t} \equiv \sigma \,
  \int\!\! \rmd \Vector{u}_{2}^{*}
  \rmd \Vector{k} \,\, 
  (\dotpr{\Vector{w^{*}}}{\Vector{k}})   \, (\hatf_{1}'\hatf_{2}' -
  \hatf_{1} \hatf_{2})
\end{equation}
To obtain an approximate solution to this equation we expand the
distribution function about a Maxwell distribution in some space as:
\begin{equation}
  \label{eq:fexpand}
  \hatf(\Vector{x},\Vector{u}) = \frac{\rho}{T_{0}} f^{0}
  [ 1 + A_{j}(\Vector{x}) \varphi_{j}(\Vector{u}) ]
\end{equation}
where, $f^{0}(\Vector{u}) \equiv {\textstyle \frac{1}{2\pi}}
\Exp{u_{i}^{*2}/T_{0}}$ is the
Maxwell distribution function with the $T_{0}$ left out of the
definition for the simplifications to follow.  The density $\rho$ is
also expanded about a leading order density field, 
\begin{equation}
  \label{eq:rhoexp}
  \rho = \rho_{0}\,(1+\rho_{1})
\end{equation}
 The functions $\varphi_{j}$ are chosen from a set
of linearly independent function space. The parameters
$A_{j}(\Vector{x})$ are determined using the method of moments. In
this method a set of functions, $\psi_{i}(\Vector{u})$, equal in
number to the number of unknowns are chosen and are multiplied with
the Boltzmann equation, and integrated over the velocity space. This
way we obtain a set of differential equations for the unknown
parameters. 

The leading order density and temperature distribution were obtained
in \cite{kum98:vib} for a dilute bed.
\begin{gather}
  \label{eq:leadrho} \rho_{0} = \frac{N\,g}{T_{0}} \, \Exp{-g\,z/T_{0}} \\
  \label{eq:leadT} T_{0} = \frac{4 \sqrt{2}}{\pi}
  \frac{\braket{U^2}}{N \sigma (1 - e^{2})}.
\end{gather}
Here, $N$ is the number of particles per unit width of the bed, $g$ is
the gravitational accelaration, $e$ is the coefficient of restitution
for particle-particle collisions, and $\braket{U^2}$ represents the
mean square velocity of the vibrating surface. For sinusoidal forcing
with characteristic velocity $U_{0}$, this is given by $\braket{U^{2}}
= U_{0}^{2}/2$.

The functions $\varphi_{j}$ are chosen from a
set of linearly independent function space. The parameters
$A_{j}(\Vector{x})$ are determined using the method of moments. In
this method a set of functions, $\psi_{i}(\Vector{u})$, equal in number
to the number of unknowns are chosen and are multiplied with the
Boltzman equation, and integrated over the velocity space. This way we
obtain a set of differential equations for the unknown parameters. 

\paragraph{Non-dimensionalisation} As a simplification we use the
following non-dimensionalisation. $\Vector{u} =
\Vector{u^{*}}/\sqrt{T_{0}}$, $z = z^{*}g/T_{0}$. Substituting
\Eqref{eq:fexpand} in \Eqref{eq:beqn}, multiplying by $\psi_{i}$ and
integrating over the velocity we obtain the steady state differential
equation for the moments for variation only in the vertical 
direction, $z$ as:
\begin{multline}
  \label{eq:lhssub}
  \frac{g}{T_{0}\sqrt{T_0}} \int \! \rmdu_{1} \,\,
    \psi_{i} u_z f^{0} \partial_{z} \; \rho ( 1 + A_{j} \varphi_{j})
    \,\, \\
  + \,\, \frac{g \rho}{T_{0}\sqrt{T_0}} \int \rmdu_{1} \,\, f^{0} ( 1 +
    A_{j} \varphi_{j}) \pder{\psi_{i}}{u_{z}} = \frac{\partial_{c}
    \hatf}{\partial t }.
\end{multline}
Here the second term has been simplified using the divergence theorem
and the condition that the distribution function vanishes for large
velocities. When the collision term is integrated over the velocities, 
$\rmdu_{1}$,
it can be simplified from the form in \Eqref{eq:colldef} to an equivalent
form (see in \cite{hcb54}, for example)
\begin{multline}
  \label{eq:collsub}
  \int \!\!\rmdu_{1} \, \psi_{i} \frac{\partial_{c} \hatf}{\partial t }  =
  \frac{1}{2} \sigma  \sqrt{T_{0}}
  \int \!\!\rmdu_{1} \rmdu_{2} \rmd \Vector{k} \,\,
  (\dotpr{\Vector{w}}{\Vector{k}})
  \rho^{2} \\
  \times f_{1}^{0} f_{2}^{0} (1 + A_{j}\varphi_{1j}) (1 +
  A_{k}\varphi_{2k})
  \,\Delta\psi_{i}
\end{multline}
where, $\Delta\psi_{i} = [\psi_{1i}' + \psi_{2i}' - \psi_{1i} -
\psi_{2i}]$, is the total change in $\psi_{i}$ due to collisions. The
terms in \Eqref{eq:lhssub} and \Eqref{eq:collsub} can be simplified by
defining $\braket{.} \equiv \int \! \rmdu \, f^{0}\,(.)$ and
$\collc[.]$ as the collision integral operator. Then we
have from \Eqref{eq:lhssub},
\begin{multline}
  \frac{g}{T_{0}\sqrt{T_0}} \bigl[ \braket{u_{z} \psi_{i}}
  \partial_{z} \rho + \braket{u_{z} \psi_{i} \varphi_{j}}
  (A_{j} \partial_{z}\rho + \rho \partial_{z} A_{j}) \bigr] \\
  + \frac{g \rho}{T_{0}\sqrt{T_0}} \left[
  \braket{\pder{\psi_{i}}{u_{z}}} + A_{j}  
  \braket{\pder{\psi_{i}}{u_{z}} \varphi_{j}} \right] \\
  = \frac{\sigma \rho^{2}}{\sqrt{T_{0}}} \bigl[ \collc[\Delta\psi_{i}]
     + A_{j}\, \collc[\Delta\psi_{i} \, \varphi_{1j}] +
     A_{k}\,\collc[\Delta\psi_{i}\, \varphi_{2k}] \\
     + A_{j}A_{k}\,\collc[\Delta\psi_{i}\, \varphi_{1j} \varphi_{2k}]
  \bigr]
\end{multline}
Further, dividing the above equation by
$\textfrac{g}{T_{0}\sqrt{T_0}}$ and defining $\rst \equiv \ln \rho$,
we obtain the following equation for the unknown variables.
\begin{multline}
  \label{eq:govmat}
  S^{0}_{i} \partial_{z}\rst + S_{ij} ( A_{j} \partial_{z}\rst +
  \partial_{z} A_{j} ) + G^{0}_{i} + G_{ij} A_{j} = \\
  \frac{\rho\, \sigma T_{0}}{g}   \left( C^{0e}_{i} + C^{0i}_{i} +
  C^{1e}_{ij} A_{j} + C^{2e}_{ijk} A_{j} A_{k} \right)
\end{multline}
where,
\begin{align*}
 S^{0}_{i} & = \braket{u_{z} \psi_{i}} \\ 
 S_{ij} & = \braket{u_{z} \psi_{i} \varphi_{j}} \\ 
 G^{0}_{i} & = \braket{\pder{\psi_{i}}{u_{z}}} \\ 
 G_{ij} & = \braket{\pder{\psi_{i}}{u_{z}} \varphi_{j}} \\ 
 C^{0e}_{i} & = \collc[\Delta\psi_{i}] \quad\text{with elastic collisions}\\
 C^{0i}_{i} & =\collc[\Delta\psi_{i}]  \quad\text{inelastic collsions,
       excluding the above term }\\
 C^{1e}_{ij} & =   \collc[\Delta\psi_{i} \, \varphi_{1j}]
         + \collc[\Delta\psi_{i}\, \varphi_{2j}]  \\
 C^{2e}_{ijk} & = \collc[\Delta\psi_{i}\, \varphi_{1j} \varphi_{2k}]
\end{align*}
Here, the superscript $e$ for the collision terms above indicate that
the collisions are considered to be elastic.

A leading order equation is obtained by considering the $A_{j}
\varphi_{j}$ as perturbations to the Maxwell distribution and by
considering elastic collision in the collision integral. Thereby we
obtain by setting $A_{j} = 0$,
\begin{equation}
  S^{0}_{i} \partial_{z} \rst + G^{0}_{i} =
  \frac{\rho\sigma T_{0}}{g} C^{0e}_{i}
    \end{equation}
With, $\psi_{i} = u_{z}$, the leading order density variation is given
by $\partial_{z} \rst = -1$, giving $\rst = -z + c$ or $\rho = \rhonot
\Exp{-z}$, where $\rhonot = Ng/T_{0} $ is the density at $z=0$.  Kumaran
\cite{kum98:vib} had obtained the values of $\rhonot$ and $T_{0}$
using a balance of the leading order source and dissipation, for low
densities are given in \Eqsref{eq:leadrho} and~\eqref{eq:leadT}. A high
density correction to these values was obtained in 
\cite{sunkum99:scal} in the leading order.  In the present analysis we
restrict ourselves to the low density limit.  To obtain a first order
balance in this limit, we neglect the quadratic term $A_{j}A_{k}$, and
subtract out the leading order equation for low densities.
\begin{equation}
  \label{eq:foeq0}
  S_{ij} \partial_{z} A_{j}  = \left[ S_{ij} - G_{ij} +
  \frac{\Exp{-z}}{\epsl} C^{1e}_{ij} \right] A_{j} +
  \frac{\Exp{-z}}{\epsl} C^{0i}
\end{equation}
Here we have used $1/\epsl \equiv \rhonot \sigma T_{0}/g = N \sigma$.
\Eqsref{eq:foeq0} is a set of coupled linear non-autonomous first order
ordinary differential equations in the varibles $A_{j}$. If we
incorporate the pertubation to density,  \Eqref{eq:rhoexp}, in \Eqref{eq:lhssub},
then the above equation for the first order quantities reads:
\begin{multline}
  \label{eq:foeq}
  S_{ij} \partial_{z} A_{j}  + S_{i}^{0}  \partial_{z} \rho_{1}
  = \left[ S_{ij} - G_{ij} + 
  \frac{\Exp{-z}}{\epsl} C^{1e}_{ij} \right] A_{j} \\
 + \left[ S_{i}^{0} - G_{i}^{0} + \frac{2\,\Exp{-z}}{\epsl} C^{0e}_{ij}
  \right]\,\rho_{1}  
  + \frac{\Exp{-z}}{\epsl} C^{0i}
\end{multline}

\subsection{Boundary conditions} The boundary conditions for the above
equations may be obtained by the following method. A balance of the
value of a moment $\phi_{i}$ is considered when it collides with a
wall moving with a velocity $U$. The change in the value of a moment
due to the collision is given by the relation:
\begin{multline}
  \Int{U}{\infty}{u_{z}^{*}}\!\!\!\Int{-\infty}{\infty}{u_{x}^{*}}
      f(\Vector{u}) 
      \phi_{i}(\Vector{u}) =  
  \Int{-\infty}{U}{u_{z}^{*}}\!\!\!\Int{-\infty}{\infty}{u_{x}^{*}}
  f(\Vector{u}) \phi_{i}(\Vector{u}) \\
  + \Int{-\infty}{U}{u_{z}^{*}}\!\!\! \Int{-\infty}{\infty}{u_{x}^{*}}
      f(\Vector{u}) \Bigl[\phi_{i}(\Vector{u}') -
      \phi_{i}(\Vector{u})\Bigr]
\end{multline}
where the primed variable deotes the velocity of the particle after a
collision.  Simplifying the above equation we get,
\begin{equation}
  \Int{-\infty}{U}{u_{z}^{*}}\!\!\! \Int{-\infty}{\infty}{u_{x}^{*}}
  f(\Vector{u}) \phi_{i}(\Vector{u}') =
  \Int{U}{\infty}{u_{z}^{*}}\!\!\!\Int{-\infty}{\infty}{u_{x}^{*}}
  f(\Vector{u}) 
      \phi_{i}(\Vector{u}) 
\end{equation}
If the mean free path of the particles is large compared to the
amplitude of vibration, then we can make an assumption that the
wall is stationary at one position, then the above equation  can be 
further simplified by averaging over different probable
velocities of the bottom wall, which, in the case of a sine wave
oscillation is,
\begin{equation}
  P(U) \rmd U = \frac{1}{\pi (U_{0}^{2} - U^{2})}
\end{equation}
if we write $U = U_{0} \sin\theta$, and $\epsilon \equiv
U_{0}^{2}/T_{0}$, then the averaged equation in terms of the
nondimensional quantities for the balance of $\phi_{i}$, after making
the substitution for the distribution function at $z=0$, will be,

\begin{widetext}
 \noindent\rule{0.5\linewidth}{1pt}\rule{1pt}{3ex}

\begin{multline}
  \label{eq:phibal}
  A_{j} \frac{1}{\pi} \Int{-\frac{\pi}{2}}{\frac{\pi}{2}}{\theta}
  \!\!\!\Int{-\infty}{\sqrt{\epsilon}\sin\theta}{u_{z}}
  \!\!\!\Int{-\infty}{\infty}{u_{x}} 
  f^{0} \phi_{i}' \varphi_{j} -
  A_{j} \frac{1}{\pi} \Int{-\frac{\pi}{2}}{\frac{\pi}{2}}{\theta}
  \!\!\!\Int{\sqrt{\epsilon}\sin\theta}{\infty}{u_{z}}
  \!\!\!\Int{-\infty}{\infty}{u_{x}} 
  f^{0} \phi_{i} \varphi_{j} \\
  =  \frac{1}{\pi} \Int{-\frac{\pi}{2}}{\frac{\pi}{2}}{\theta}
  \!\!\!\Int{\sqrt{\epsilon}\sin\theta}{\infty}{u_{z}}
  \!\!\!\Int{-\infty}{\infty}{u_{x}} 
  f^{0} \phi_{i} - 
   \frac{1}{\pi} \Int{-\frac{\pi}{2}}{\frac{\pi}{2}}{\theta}
  \!\!\!\Int{\infty}{\sqrt{\epsilon}\sin\theta}{u_{z}}
  \!\!\!\Int{-\infty}{\infty}{u_{x}} 
  f^{0} \phi_{i}' 
\end{multline}
\noindent\hspace*{0.5\linewidth}
\rule[-3ex]{1pt}{3ex}\hspace{-1pt}\rule{0.495\linewidth}{1pt}
\end{widetext}
or simply as,
\begin{equation}
  \label{eq:bceq}
  E_{ij} A_{j}(0) = B_{i}
\end{equation}  
Solving these simultaneous equations we obtain conditions satisfied by
$A_{j}$ at $z=0$.

\section{Third-order moment formulation} \label{sec:third}

We now take the specific case of the third order formulation. A third
order formulation is the lowest order in which we can include the
anisotropy in the distribution function.  We now briefly explain the
choice of the moments $\varphi_{j}$. The distribution function
satisfies the following criteria. (i) the time averaged vertical flux
is zero, $\braket{u_{z}} = 0$, (ii) the distribution function is
normalised to unity, (iii) it is symmetric in the horizontal
velocities. It is convenient to choose the lower order moments from a
multi-dimensional Hermite polynomials for the following reasons. These
polynomials form a linearly independent orthogonal basis, and the
resulting equations are more convenient than a linearly independent
set. In addition two of the above conditions will be automatically
satisfied by the distribution function by setting the corresponding
the corresponding $A_{j}$ to zero. Note that since the leading order
distribution function $f^{0}$ already satisfies the unit normalisation
condition, we would require that $A_{j}\braket{\varphi_{j}} = 0$. The
flux condition requires that $A_{j} \braket{u_{z}\varphi_{j}} = 0$.
Since $1$ and $u_{z}$ are two of the linearly independent functions in
the orthogonal Hermite polynomials, setting the coefficients of these
two will ensure that the above two conditions will be satisfied at all
orders of the polynomials. Thus the choice of the $\varphi_{j}$
becomes simple by incorporating the constraints on the distribution
function directly.

The set of multi-dimensional orthogonal polynomials can be obtained
from the recurrence relation for the Hermite polynomials,
\begin{gather}
H^{n+1}(x) = x H^{n}(x) - n H^{n-1}(x) \\
\text{with} \quad H^{0}(x) = 1, \quad H^{1}(x) = x \nonumber \\
\intertext{and the multidimensional set is then given by,}
H(u_{x},u_{z}) =  H^{m}(u_{x}) H^{n}(u_{z}) \quad \text{ even } m,
\text{ all } n.
\end{gather}
A symmetric distribution in the horizontal velocity can be ensured by
taking only even powers of $u_{x}$, ie., even $m$ in the above
expression. The above polynomials are can be normalised to unity and
the factor is $\textfrac{1}{m! \, n!}$.

In the case of the third order approximation we obtain the following
polynomials,
\begin{equation}
  \label{eq:vphiset}
  \varphi_{j} = \{ -1 + u_z^2 , -3 u_z + u_z^{3} , -1 +
  u_x^{2} , -u_z + u_x^{2}  u_z \},
\end{equation}
in which the first two members viz., $1$ and $u_{z}$ have been
omitted for reasons discussed above. We choose the same set for the
moment generating functions 
$\psi_{i}$ in \Eqref{eq:foeq}  and the functions for the boundary
conditions, $\phi_{i}$ in \Eqref{eq:bceq}. This way we can get the same
order of representation in the moment equations as well as in the boundary
conditions. 

The moments of the distribution can be obtained by substituting the
expression~\eqref{eq:vphiset} in \Eqref{eq:fexpand}, which are:
\begin{subequations}
\label{eq:moms}
\begin{align}
  \label{eq:ux2}
  \braket{u_{x}^{2}} &= (1 + 2 A_{3}) \\
  \label{eq:uz2}
  \braket{u_{z}^{2}} &= (1 + 2 A_{1}) \\
  \label{eq:ux2uz}
  \braket{u_{x}^{2}u_{z}} &= 2\,  A_{4} \\
  \label{eq:uz3}
  \braket{u_{z}^{3}} &= 6\,A_{2} 
\end{align}
\end{subequations}

\section{Solution}
With the moments considered in \Eqref{eq:vphiset}, the set of relations
in \Eqref{eq:foeq0} for $A_{i}$ can be written as
\begin{equation}
  \partial_{z} A_{i} = L^{0}_{ij} A_{j} + L^{1}_{ij} A_{j} \Exp{-z} +
  b^{1}_{i} \Exp{-z},
\end{equation}
where,
\begin{equation*}
  L^{0}_{ij} = S_{ik}^{-1}(S_{kj} - G_{kj}), \quad
  L^{1}_{ij} = \frac{1}{\epsl} S_{ik}^{-1}C^{1e}_{kj}, \quad
  b^{1}_{i}  = \frac{1}{\epsl} S_{ik}^{-1}C^{0i}_{k},
\end{equation*}
and by considering a moment $\psi_{i} = u_{z}$ in \Eqref{eq:foeq} we
have for the density correction, 
\begin{equation}
  \label{eq:rho1de}
  \partial_{z}\rho_{1} + 2\,\partial_{z}A_{1} = 2\,A_{1},
\end{equation}
In the case of the third order approximation, we have
\begin{align*}
  L^{0}_{ij} &=  \{ \{ 0,0,0,0\} ,\{ 0,1,0,0\} ,\{ 0,0,0,0\} ,\{ 0,0,0,1\} \} \\
  L^{1}_{ij} &= \textstyle \frac{\sqrt{\pi}} {\epsl} \{ \{
  0,-{\frac{3}{2}},0,{\frac{1}{2}}\} , 
  \{ -{\frac{1}{3}},0,{\frac{1}{3}},0\} ,
  \{ 0,{\frac{3}{2}},0,-{\frac{5}{2}}\},\\
  & \qquad \{ 1,0,-1,0\} \} \\
  b^{1}_{i} &= \textstyle - \frac{(1-e^{2}) \sqrt{\pi}}{6 \epsl}
  \{ 0,1,0,3\} 
\end{align*}
We note here that the $A_{i}$ are independent of the density
correction $\rho_{1}$ in the first order approximation 
These equations can be rearranged into a single fourth order equation
in $A_{1}$, which is easily accomplished through a symbolic routine:
\begin{multline}
  \label{eq:nthord}
  A_1^{(4)}(z) = - 4\,A_1^{(3)}(z) + \left( \frac{4\pi}{\epsl^{2}}
    \Exp{-2z} - 4\right) A_{1}''(z) \\
  - \frac{\pi^{2}}{\epsl^{4}}(1-e^{2}) \,\Exp{-4z}
\end{multline}
With this simplification, the other variables can be written down in
terms of $A_{1}$ and its derivatives as
\begin{subequations}
\label{eq:otherfns}
\begin{multline}
  A_{2}(z) = \frac{\left( -1 + {e^2} \right)} {24} \ezfact^{-1} \\
  + \left(-A_{1}'(z) + \frac{A_{1}''(z)}{6} \right) \, \ezfact \, \\
  + \left( \frac{A_{1}''(z)}{6} - \frac{A_{1}^{(4)}(z)}{24} \right) 
  \ezfact^{3}
\end{multline} 
\begin{equation}
  A_{3}(z) =   A_{1}(z) - A_{1}''(z) \ezfact^{2}
\end{equation}
\begin{multline}
  A_{4}(z) =  \frac{\left( -1 + {e^2} \right)}{8} \, \ezfact^{-1} \\
  + \left(-A_{1}'(z) + \frac{A_{1}''(z)}{2} \right) \, \ezfact \\
  + \frac{1}{2}\, \left( A_{1}''(z) - \frac{A_{1}^{(4)}(z)}{4} \right) 
  \, \ezfact^{3}
\end{multline}
\end{subequations}

To solve \Eqref{eq:nthord}, we can make a reduction in order by the
following substitution
\begin{equation}
  \label{eq:a2y}
  A''(z) =  y(z) 
\end{equation}
With the transformation $x=\Exp{-z}$, we obtain
from \Eqref{eq:nthord}, (Note: here $x$ is just a transformation variable
and not the cartesian co-ordinate)
\begin{equation}
  \label{eq:ydeq}
  \ddot{y} - 3 \frac{\dot{y}}{x} - \frac{1}{x^{2}} (c\,x^{2} - 4)\,y =
  D \, x^{2}
\end{equation}
where a dot accent denotes $\partial_{x}$, and the constants are $c
\equiv \textfrac{4\pi}{\epsl^{2}}$, $ D \equiv -\textfrac{\pi
  (1-e^{2})}{\epsl^{4}}$. Further with
\begin{equation}
  \label{eq:y2w}
  w(x) = y(x)/x^{2},
\end{equation}
we get,
\begin{equation}
  \label{eq:wdeq}
  \ddot{w} +  \frac{\dot{w}}{x} - c\,w = D, 
\end{equation}
which is a modified Bessel equation of zeroth order. The solution to the
homogeneous equation is given by:
\begin{equation}
  \label{eq:wh}
  w_{h}(x) = c_{1}\, I_{0}(\scx) + c_{2}\, K_{0}(\scx) 
\end{equation}
A particular solution to the inhomogeneous equation can be obtained by
variation of parameters. The Wronskian of the above solutions
can be written down from standard references,
\begin{equation}
  \begin{vmatrix}
    I_{0}(\scx) &  K_{0}(\scx) \\ 
    \dot{I_{0}}(\scx) &  \dot{K_{0}}(\scx)
  \end{vmatrix} = -\frac{1}{x}.
\end{equation}
The particular solution so obtained is given by:
\begin{equation}
  \label{eq:wp}
  w_{p}(x) = - \frac{D \, x}{\sqrt{c}} \, \left[ I_{0}(\scx) \,
  K_{1}(\scx) + K_{0}(\scx) \, I_{1}(\scx) \right].
\end{equation}
The most general solution to the inhomogeneous equation is then
\begin{equation}
  \label{eq:wgen}
  w(x) = w_{h} + w_{p}(x).
\end{equation}
The \Eqref{eq:a2y} can now be solved in terms of the independent variable
$x$, by further reduction in order, by writing it, 
\begin{gather}
  \label{eq:adot}
  x \, (x \ddot{A}_{1}(x) + \dot{A}_{1}(x) ) = y(x),
\intertext{as:}
  \label{eq:bdeq}
  \dot{B}(x) = w(x) - \frac{B(x)}{x}
\intertext{where,}
  \label{eq:a2b}
  \dot{A_{1}}(x) = B(x).
\end{gather}
The solutions to these first order equations can be easily written as
\begin{equation}
  \label{eq:bsol}
  B(x) = \frac{1}{x} \, \Biggl[ \Int{}{x}{x'} w(x') \, x' \enspace +
  \enspace c_{3} \Biggr]
\end{equation}
and
\begin{equation}
  \label{eq:asol}
  A_{1}(x) = \Int{}{x}{x''} \frac{1}{x''} \, \Biggl[ \Int{}{x''}{x'} w(x') \, x'
  \enspace + \enspace c_{3} \Biggr] \enspace + \enspace c_{4}.
\end{equation}
The correction to the density is then from \Eqref{eq:rho1de},
\begin{equation}
  \label{eq:rhosol}
  \rho_{1}(x) = -2 \Int{}{x}{x'} \left(\frac{A_{1}(x')}{x'} + B(x')
  \right) + c_{5}.
\end{equation}
The derivatives of $A_{1}(z)$ can be written down as
\begin{subequations}
\label{eq:a1ders}
\begin{align}
  A_{1}'(z) & =  -x\, \dot{A_{1}}(x) = -x\,B(x) \\
  A_{1}''(z) & =  -x^{2} \, w(x) \\
  A_{1}'''(z) & =  -2x^{2} \, w(x) -x^{3}\,\dot{w}(x)
\end{align}
\end{subequations}
These are used to evaluate the other unknown functions in
\Eqsref{eq:otherfns}.

\subsection{Boundary conditions}
Some of the constants of integration can be directly eliminated by
requiring that the functions to take finite values for $x=0$ (or
equivalently, $z\rightarrow \infty$). Consider one of the linearly
independent solutions of $w_{h}(x)$ in \Eqref{eq:wh}, $c_{2}\,
K_{0}(\scx)$, which when integrated through \Eqsref{eq:bsol}
and~\eqref{eq:asol} gives a term in $A_{1}(x)$: $c_{2}\,
K_{0}(\scx)/c$. Since $K_{0}(\scx)$ is singular as $x\rightarrow 0$,
we require $c_{2} = 0 $.  Similarly we require that $c_{3} = 0$, as
this will lead to a singular term in $A_{1}(x)$: $c_{3}\,\ln x$.

The expressions in \Eqref{eq:wp} cannot be integrated to obtain a
closed form expression. Nevertheless, they can be easily evaluated
numerically by converting the expressions \Eqref{eq:bsol}
and~\eqref{eq:asol} to a definite integral, $\tInt{x_{0}}{x}{x'}$
added with some constant. The choice of the lower limit, $x_{0}$, can,
in general, be artibary to suit the convinience of matching the given
boundary conditions; but such a choice, say $\tInt{1}{x}{x'}$, for the
integral in \Eqref{eq:bsol} will lead to singularities in the
expression for $A_{1}(x)$ in \Eqref{eq:asol}, similar to those
obtained by the constant $c_{3}$ term. To do away with this
singularity, we choose $x_{0}=0$, expand the integral in
\Eqref{eq:bsol} in Taylor's series, and subtract out the source of the
singularity (which essentially is the constant of integration, \ie,
the value of the integral evaluated at the lower limit).  In expanding
$w(x)$ about $x=0$, we note from \Eqref{eq:wgen} that $w(0) = 0$,
therefore,
\begin{align*}
B(x) =   \frac{1}{x} \Int{0}{x}{x'} w(x')\,x' 
  &= \frac{1}{x} \Int{0}{x}{x'}  \left(w_{1}x' + w_{2} x^{\prime2} +
    \ldots \right) \, x' \\
  &=  \left(\frac{w_{1}}{2} x + \frac{w_{2}}{3} x^{2} + \ldots
  \right),
\end{align*}
where $w_{1}, w_{2}, \ldots$ are constants coming from the Taylor's
expansion. Such an expansion is possible as the series converges for
$0\le x <1$. The above expression identically vanishes at $x=0$, therefore we
donot have to explicitly subtract out any singularity during numerical
computation of the definite integral. Such problems of subtracting out
the singularity, however, does not arise in the integral of
\Eqref{eq:asol} and any arbitary point, $x_{0}$, can simply be chosen for
the numerical integration, keeping in mind the boundary conditions.

We are now left with two arbitary constants, $c_{1}$ and $c_{4}$.
These are evaluated using boundary conditions in a manner similar to
those used in \cite{kum98:vib}. For the sake of convinience we choose
the lower limit for the definite integral in \Eqref{eq:asol} to be $x_{0}
= 1$, then $c_{4}$ will simply be equal to the value of $A_{1}(x=1)$.
It can be seen from \Eqref{eq:uz2} that $A_{1}$ is directly propotional
to the moment of the distribution function $\braket{u_{z}^{2}}$, whose
value at $z=0$ can be obtained by considering it as a vertical flux of
momentum because of collisions with the wall. Assuming that the
particles collide with the wall have a leading distribution to be a
Maxwellian, the flux of momentum along the vertical direction is given
by,
\begin{equation}
  \label{eq:bc1}
  \evalat{\braket{u_{z}^{2}}}{z=0}  = 1 + \frac{\epsilon}{2}
\end{equation}
substituting in  \Eqref{eq:uz2} we have, $\evalat{A_{1}}{(x=1)} =  \epsilon/4$,
therefore,
\begin{equation}
  c_{4} = \frac{\epsilon}{4}.
\end{equation}
Since, momentum is transferred only in the vertical direction we have
from \Eqsref{eq:moms}, $A_{3}\vert_{x=1} = 0$ and $A_{4}\vert_{x=1} = 0$.
The constant, $c_{1}$, can now be evaluated from \Eqsref{eq:a2y},
\eqref{eq:y2w} and~\eqref{eq:wgen}.
\begin{equation}
  \label{eq:c1}
  c_{1} = \frac{1}{I_{0}(\sqrt{c})} \, \evalat{\left(A_{1}''(z) -
  w_{p}(x) \right)}{z=0\text{ or } x=1},
\end{equation}
The constant $\evalat{A_{1}''(z)}{z=0}$ can be solved for from
\Eqsref{eq:otherfns}
\begin{align*}
  \evalat{A_{1}''(z)}{z=0} &= \frac{\pi}{\epsl^{2}}\,
  \evalat{\left( A_{1} + A_{3} \right)}{z=0} \\
  &= \frac{\pi\,\epsilon}{4\,\epsl^{2}}
\end{align*}
The other constant $c_{5}$, is obtained by considering mass balance for
the density correction $\rho_{1}$. Since the total mass of the bed is
balanced in the leading order density profile $\rho_{0}$, the
balance for the correction to the density is given by
\begin{gather}
  \label{eq:rhobc}
  \Int{0}{\infty}{z^{*}} \rho_{0} \rho_{1} = 0, \\
\intertext{or,} 
  \Int{0}{1}{x} \rho_{1} = 0.
\end{gather}

We note here that we have not strictly incorporated the boundary
conditions as discussed in \secref{sec:form}. This is because of a need to
satisfy more strong boundary conditions of non-diverging solutions for
$z\rightarrow\infty$. The equations of the sort of \Eqref{eq:bceq} would be
useful in obtaining, for example, a series solution by numerical
methods where we expand the solution in decaying functions such as
Laguerre polynomials.

Furthermore, the moments related to the horizontal direction are not
exactly satisfied,  \ie, the boundary conditions related to $A_{3}$ and
$A_{4}$. While the value of $A_{3}$ is not satisfied independently,
but only partly in \Eqref{eq:c1}, $A_{4}$ is never used. These can be used
only in a higher order polynomial approximation. Strictly  therefore,
only $A_{1}$ is satisfied exactly.


\section{Discussion}
The results obtained here are qualitatively the same as obtained in
\cite{kum98:vib}. At the time of the previous work, the only results
available were some experimental measurements of \cite{warretal95} and
only a qualitative comparison was possible to ascertain the validity
of the theory. We have now compared the results of with a numerical
Event Driven (ED) simulation of vibrated hard disks. The ED simulation
was done with periodic boundaries and with an approximate
representation of the bottom wall \cite{sunkum:mom}.  The following
figures show that the theory is indeed good in the limit of its
validity. The figures also show a comparison with the approximate
series solution which was obtained in \cite{kum98:vib}. In most cases
the series solution provides a good approximate.

\begin{figure}[tp]
  \begin{center}
    \xlabel{$z/\sigma$}
    \ylabel{$\nu$}
    \gpfig{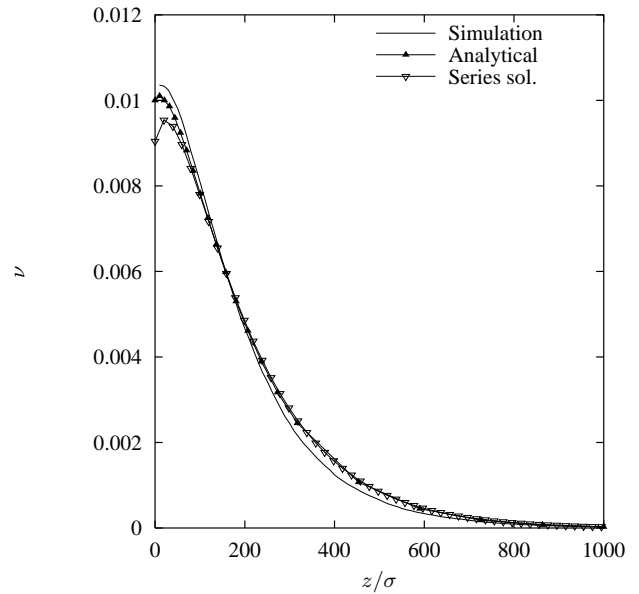} 
    \caption{Density profile for $\ndp = 3$, $\epsilon=0.3$. The
    analytical solution obtained in this paper and the series solution
    of \cite{kum98:vib} show good agreement. There is a negative
    correction to the density at the bottom after which 
    it decays exponentially as the leading order values.}
    \label{fig:dens}
  \end{center}
\end{figure}

The parameters chosen for the simulation are in correspondence with
the limit of validity of the theory: $\epsilon \ll 1$, and $\ndp \sim
\OrderOf{1}$, (see \cite{sunkum99:scal,sunkum:mom}, for a discussion).
There is a small negative correction to the density near the bottom
wall, as shown in \Figref{fig:dens}, due to the energy flow near the
bottom wall, after which it falls of exponentially as the leading
order density.

One important difference, in the formulation of the density, between
the present and the previous work \cite{kum98:vib} is the following. In
the previous work a correction to the distribution function due to
variations in the distribution function over distances comparable to
the particle diameter were described by using a small parameter,
$\epsilon_{G}$.  It was shown in a later work \cite{sunkum:mom}
that this correction is essentially equivalent to the high density
correction obtained on the lines of Enskog correction to dense gases,
therefore it has been omitted in the present consideration of dilute
bed.

\begin{figure}[tp]
  \begin{center}
    \xlabel{$z/\sigma$}
    \ylabel{$\braket{u_{x}^{2}}$}
    \gpfig{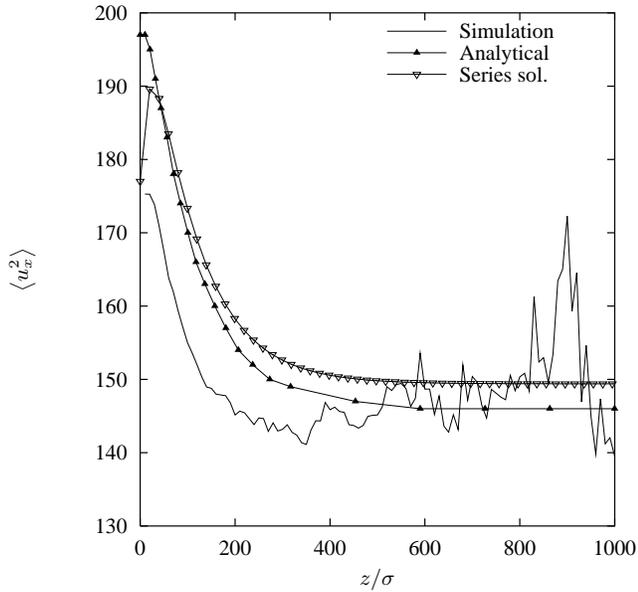}
    \caption{Horizontal temperature profile for $\ndp = 3$,
      $\epsilon=0.3$. A magnitude comparison with \Figref{fig:uz2} shows
      the degree of anisotropy in a vibrated bed.}
    \label{fig:ux2}
  \end{center}
\end{figure}

\begin{figure}[tp]
  \begin{center}
    \xlabel{$z/\sigma$}
    \ylabel{$\braket{u_{z}^{2}}$}
    \gpfig{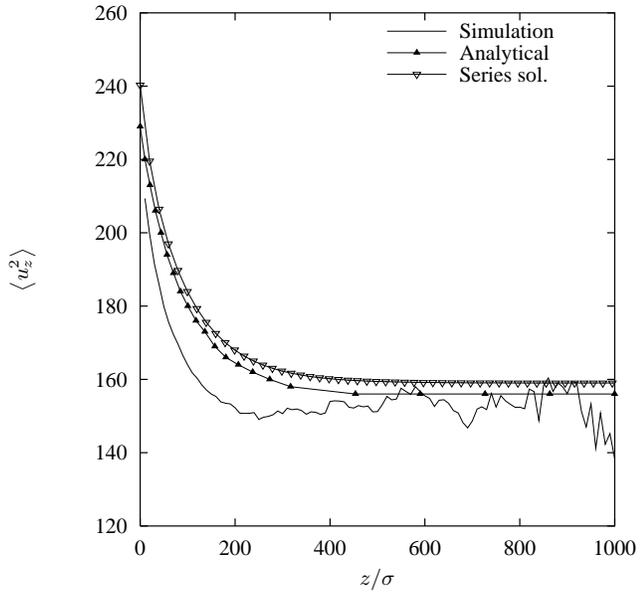}
    \caption{Vertical temperature profile for $\ndp = 3$,
    $\epsilon=0.3$. The theoretical values are closer to the
    simulation here, than for the horizontal temperature because of
    the boundary conditions imposed.}
    \label{fig:uz2}
  \end{center}
\end{figure}

\begin{figure}[tp]
  \begin{center}
    \xlabel{$z/\sigma$}
    \ylabel{$\braket{u_{z}^{3}}$}
    \gpfig{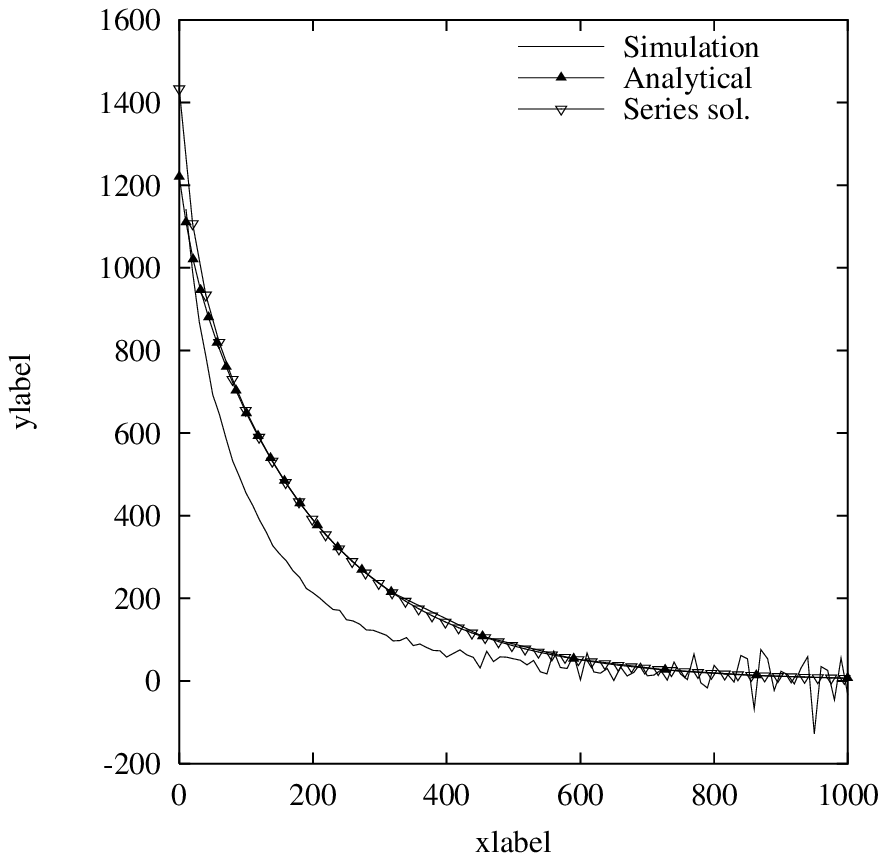}
    \caption{Flux of energy in the vertical direction for $\ndp = 3$,
      $\epsilon=0.3$. Here again, the theoretical values compare well
      in order of magnitude, because of boundary conditions are
      imposed in this direction.}
    \label{fig:uz3}
  \end{center}
\end{figure}

Due to the anisotropic nature of the energy input to the
vibro-fluidised bed, the lower order moments of the distribution
clearly show the anisotropy. \Figsref{fig:ux2} and~\ref{fig:uz2} show the
horizontal and vertical temperature profiles, while \Figref{fig:uz3} shows
the vertical flow of energy. Anisotropies were also observed by us in
deep bed simulations of disks which had the wave-like surface patterns
\cite{sun:wavesim}; although the nature of anisotropy was more
pronounced even in the shape of the distribution function itself (the
vertical distribution had double peaks and the horizontal distribution
had single peak and exponential tails). Could the presence of
anisotropy be an important feature giving rise to an instability in
one direction ? A stability analysis of the solution from the present
analysis model might help resolve this. The usual models based on
hydrodynamic equations do not take into account this anisotropy.

\begin{figure}[tp]
  \begin{center}
    \xlabel{$z/\sigma$}
    \ylabel{$\braket{u_{z}^{2}}$}
    \gpfig{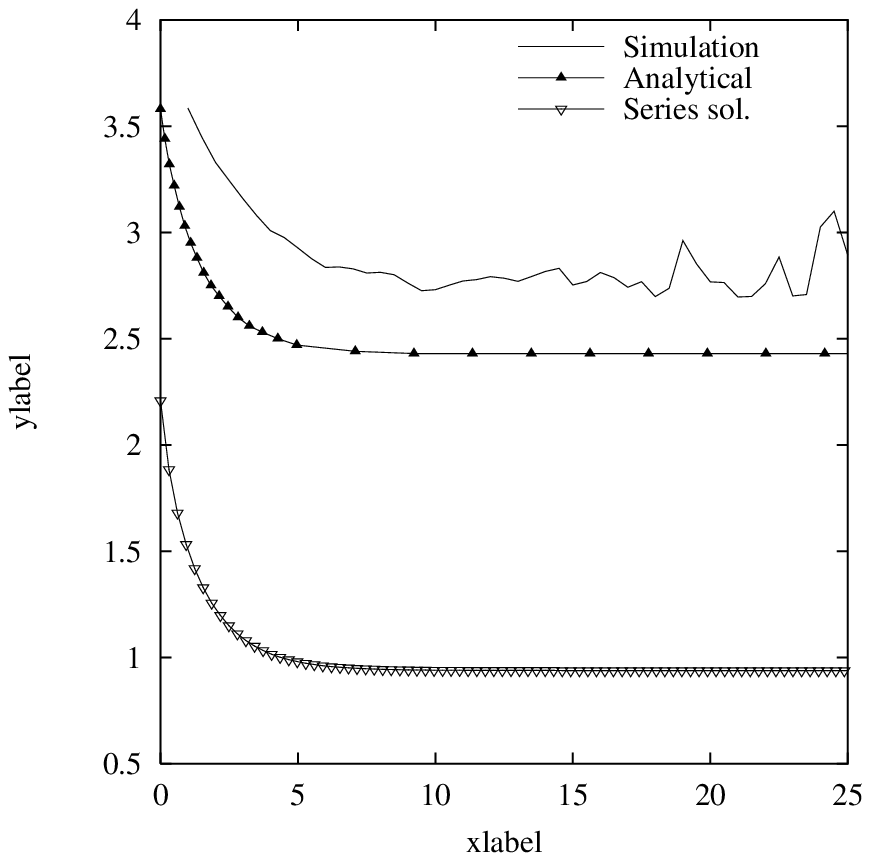}
    \caption{Vertical temperature profile prediction is good even when
      the density is high ($\nu \sim 0.3i$).}
    \label{fig:uz2d}
  \end{center}
\end{figure}

\begin{figure}[tp]
  \begin{center}
    \xlabel{$z/\sigma$}
    \ylabel{$\braket{u_{z}^{3}}$}
    \gpfig{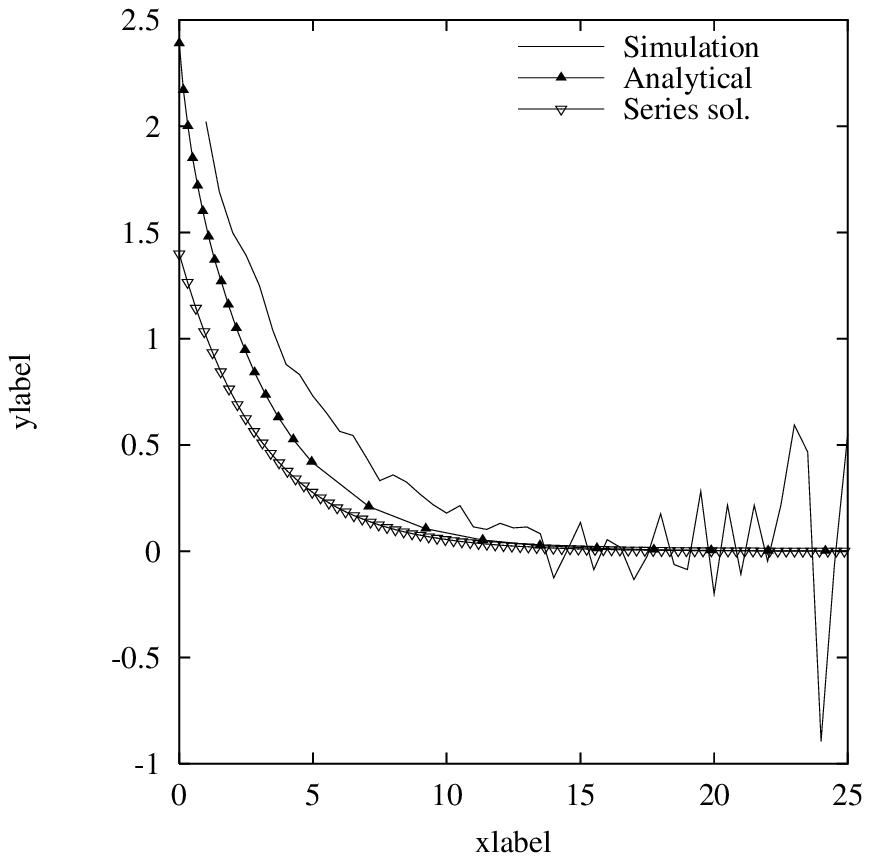}
    \caption{Flux of energy prediction is good even when the density
      is high ($\nu \sim 0.3 $).}
    \label{fig:uz3d}
  \end{center}
\end{figure}

As pointed out earlier, the density correction considered here is
different from the one used in the previous work, in that the high
density correction is not considered in first order in the
distribution function.  This results in
a better prediction of the vertical temperature \Figref{fig:uz2d} and
flux of energy \Figref{fig:uz3d} even when the density prediction is not
expected to be good.

To conclude, we have shown that (a) It is possible to obtain an
analytical solution to the Boltzmann equation for vibro fluidised bed
in the low density limit correct upto the third order in the moments
of the distribution function. (b) The qualitative nature of results
obtained here are similar to those obtained in \cite{kum98:vib}. (c)
The correction to the distribution function due to spatial variation
of the order of a particle diameter was neglected here as this is
turns out to be a a correction to a higher order in density
\cite{sunkum:mom}. With this it is seen that the density still shows a
negative correction at the bottom wall due to the energy flow. (d) The
boundary conditions are overspecified in the problem and we chose to
satisfy exactly, only the ones involved in the momentum transfer the vertical
direction. (e) Even with this restricted choice, the theoretical
values for the different moments 
compare reasonably well with the simulation particularly in the
anisotropy exhibited.  (f) This gives us some 
confidence to explore further with the stability of the solution, and
with an higher order approximation to the distribution function if
required.


\end{document}